\documentclass[aps,prd,preprint,tightenlines,groupedaddress,showpacs]{revtex4}
\usepackage{epsf,epsfig,graphics,graphicx}
\begin{document}

\title{Nonequilibrium Effects and Baryogenesis}

\author{Yeo-Yie Charng$^1$}
\email{charng@phys.sinica.edu.tw}
\author{Da-Shin Lee$^{2,3}$}
\email{dslee@mail.ndhu.edu.tw}\thanks{corresponding author.}
 \author{Chung Ngoc Leung$^4$}
 \email{leung@physics.udel.edu}
\author{Kin-Wang Ng$^1$}
\email{nkw@phys.sinica.edu.tw}
\affiliation{$^1$Institute of
Physics, Academia Sinica, Taipei,
Taiwan 115, R.O.C.\\
$^2$Department of Physics, National Dong Hwa University,
Hua-Lien, Taiwan 974, R.O.C.\\
$^3$The Theory Group, Blackett Laboratory,Imperial College, London
SW7 2BZ, U.K. \\
$^4$Department of Physics and Astronomy, University of Delaware,
Newark, Delaware 19716, U.S.A.}

\vspace*{0.6 cm}
\date{\today}

\vspace*{1.2 cm}

\begin{abstract}
Possible effects due to nonequilibrium dynamics in the
Affleck-Dine mechanism of baryogenesis are examined. Using the
closed-time-path formalism, the quantum fluctuation and the
back reaction of the Affleck-Dine scalar field are incorporated
self-consistently into the dynamical equations of the system by
invoking a nonperturbative Hartree approximation. It is found
that such nonequilibrium effects can significantly affect the
amount of baryon asymmetry that can be generated. In particular,
it is possible to generate the observed baryon asymmetry with
suitable initial conditions. The methodology described in this
paper as well as some of the results obtained are quite general,
and can be applied to any complex scalar field in a cosmological
background.

\end{abstract}

\maketitle

\section{Introduction}

It is evident that no concentration of antimatter exists within
the Solar system and the Milky Way. The absence of annihilation
radiation from the Virgo cluster indicates that antimatter can
hardly be found within a 20 Mpc scale. A study of the contribution
of annihilation radiation near the matter-antimatter boundaries to
the cosmic diffuse gamma-ray background virtually excludes domains
of antimatter in the visible Universe~\cite{cohen}.

Direct observations of luminous matter show that baryons constitute
about a few percents of the total mass of the Universe. This gives
a value of order $10^{-10}$ for $n_{\rm B}/s$, the ratio of the 
baryon number density to the entropy density. Precise measurements 
of the abundance of primordial light elements predicted in big-bang
nucleosynthesis combined with the cosmic microwave background 
experiments restrict this ratio in the range 
$n_{\rm B}/s\simeq (5-10)\times 10^{-11}$~\cite{fields}.

Many scenarios for baryon production in the early Universe have
been proposed to explain this small baryon asymmetry, such as
baryogenesis in grand unified theories, electroweak baryogenesis,
leptogenesis, and Affleck-Dine (AD) baryogensis~\cite{enqvist}.
Affleck and Dine~\cite{afdine} proposed a mechanism of baryogensis
in supersymmetric models in which scalar quark and lepton fields
obtain large vacuum expectation values along flat directions of
the scalar potential.  These coherent scalars or the condensate
start to oscillate at a temperature of order 10~TeV when
supersymmetry-breaking effects start to become important, and a
net baryon number is developed and stored in the oscillating
fields provided that $C$ and $CP$ symmetries are also violated.
Subsequently, the scalar quark and lepton fields decay and produce
a baryon asymmetry.  In fact, the AD mechanism is too efficient
and the resulting baryon asymmetry is usually too large. Several
dilution processes have been considered to reduce the large baryon
asymmetry to the observed value.  They involve either introducing
additional entropy releases after baryogenesis (e.g., by the decays
of the inflaton~\cite{Linde}, massive field fluctuations~\cite{kkk}, 
or the dilaton~\cite{dolgov}) or reducing the baryon production by 
invoking nonrenormalizable terms~\cite{ng}.  In all of these studies, 
the scalar fermion fields have been treated as classical fields and 
obey the classical equations of motion. In this paper, we will be 
concerned with quantum fluctuations created from the dynamics of 
the Affleck-Dine scalar field and its back reaction effects.  In 
particular, we study how these nonequilibrium processes affect the 
baryon production in the AD mechanism.

It has been proposed that in a certain AD flat direction the
condensate is not the state of lowest energy but fragments to 
form metastable or stable Q-balls, depending on the shape of the
radiative correction to the flat direction~\cite{enqvist}.  For
instance, this occurs along the flat direction with large mixtures
of scalar top quark and light gaugino masses or in the 
$H_u L$-direction.  In this paper we estimate the baryon asymmetry 
assuming that the AD condensate does not lead to this type of Q-ball 
formation.  Our presentation is organized as follows.  The model of 
Affleck-Dine baryogenesis is described in Sec.~II.  We introduce the 
method of nonequilibrium field theory in Sec.~III.  It is then applied 
to the Affleck-Dine model in Sec.~IV.  We show in Sec.~V how to
incorporate nonequilibrium effects in the calculation of the baryon
asymmetry, and present the results of our numerical calculation and
discuss their implications in Sec.~VI.  Our conclusions are offered
in Sec.~VII.

\section{Affleck-Dine Baryogenesis}

Affleck and Dine~\cite{afdine} have shown that, in a supersymmetric 
SU(5) grand unified model, there is a flat direction in the low-energy 
effective potential for the following set of vacuum states of the up 
($\tilde{u}$), strange ($\tilde{s}$), and bottom ($\tilde{b}$) squark 
fields as well as the scalar muon field
($\tilde{\mu}$):
\begin{equation}
 \left<\tilde{u}_3^c\right>=a,\;\left<\tilde{u}_1\right>=v;\quad
 \left<\tilde{s}_2^c\right>=a,\;\left<\tilde{\mu}\right>=v;\quad
 \left<\tilde{b}_1^c\right>=e^{i\xi}\sqrt{|v|^2+|a|^2},
\end{equation}
with all other fields having vanishing vacuum expectation values.
Here the superscript $c$ denotes charge conjugation, the subscripts
denote color indices, $a$ and $v$ are arbitrary complex numbers,
and $\xi$ is real.  A large baryon asymmetry can be generated by the
decays of the condensate associated with these vacuum expectation
values via baryon-number violating dimension-four scalar couplings
which appear after supersymmetry has been broken.  To illustrate how
the mechanism works, Affleck and Dine considered a toy model with a
single complex scalar field $\Phi$ described by the Lagrangian density
\begin{equation}
 {\cal L} = \left(\partial_{\mu} \Phi^{\dagger} \right) \left(
 \partial^{\mu} \Phi \right)- m_\Phi^2 \Phi^\dagger \Phi -
 i \lambda \left(\Phi^4 - \Phi^{\dagger\,4} \right),
\label{model}
\end{equation}
which contains the mass term with $m_\Phi^2 = M_S^2$ and the
baryon-number violating coupling, $\lambda \sim \epsilon M_S^2/{M_G^2}$,
where $M_S$ is the effective supersymmetry breaking scale, $\epsilon$
is a real parameter characterizing $CP$ violation, and $M_G$ is the
grand unification scale.  For small $\Phi$, the theory has an
approximately conserved current
\begin{equation}
 j_\mu = i \left( \Phi^\dagger \partial_\mu \Phi - (\partial_\mu
 \Phi^\dagger) \Phi \right)
\label{Bcurrent}
\end{equation}
due to the approximate global $U(1)$ symmetry: $\Phi \rightarrow
e^{i \theta} \Phi$.  The corresponding charge will be referred to as
the baryon number (this will be the case if, for example, $\Phi$
represents a scalar quark field).

Classically, in an expanding universe, $\Phi$ obeys the equation
of motion
\begin{equation}
 \frac{d^2\Phi}{dt^2}+3H\frac{d\Phi}{dt}+m_\Phi^2\Phi =
 4i\lambda\Phi^{\dagger\,3},
\end{equation}
where $H$ is the Hubble parameter. With the initial conditions
at $t = t_0$:
\begin{equation}
 \Phi\vert_{t=t_0} = i\Phi_0
 \quad {\rm and} \quad
 \dot{\Phi}\vert_{t=t_0} = 0,
\label{initialcond}
\end{equation}
where $\Phi_0$ is real and $\dot{\Phi} ={d\Phi}/{dt}$, it
was found that the baryon number per particle at large
times ($t \gg m_\Phi^{-1}$) in either a matter-dominated or a
radiation-dominated universe is given by
\begin{equation}
 r \simeq \lambda \Phi_0^2/m_\Phi^2.
\label{rAD}
\end{equation}
Assuming $m_\Phi^2 = M_S^2$ and $\lambda = \epsilon
M_S^2/{M_G^2}$, we have $r \simeq \epsilon \Phi_0^2/M_G^2$, which
can easily provide a large initial $n_{\rm B}/s$ and thus dilution
processes have to be introduced to reduce it to the observed
value. For example, taking $\epsilon=10^{-3}$, $M_S=10^{-16}M_P$,
$M_G=10^{-1}M_P$, and $\Phi_0^2 = 10^{-3} M_P^2$, where $M_P$ is
the Planck mass, we find $\lambda=10^{-33}$ and $r \simeq
10^{-4}$. Note that the decay width of the condensate can be
estimated as $\Gamma_\Phi\sim (\alpha_s/\pi)^2
m_\Phi^3/|\Phi|^2$~\cite{afdine}, which is typically much smaller
than the frequency of the oscillating scalar fields, i.e.,
$\Gamma_\Phi\ll m_\Phi$. Therefore, a net baryon number is
developed and gets saturated in the oscillating scalar quark and
lepton fields before they decay and produce a baryon asymmetry.

The above consideration did not take into account the fact that a
dynamical background field could generate quantum fluctuations
which could in turn influence its evolution~\cite{boyan1,hu,
mottola,lee}.  Substantial quantum fluctuations from the vacuum
state may lead to a value of the generated baryon number density
quite different from that obtained from purely classical
arguments. In the following sections, we will study how the
quantum fluctuation of the dynamical $\Phi$ field and the back
reaction on its evolution will affect the ratio $r$ for the model
of Eq.~(\ref{model}). Nonperturbative Hartree factorizations will
be implemented along with the method of nonequilibrium quantum
field theory to self-consistently take account of quantum
fluctuations and back-reaction effects on the background field. To
be completely general, we will not fix the values of $m_\Phi$,
$\lambda$, and $\Phi_0$, leaving them as free parameters.

\section{Nonequilibrium Quantum Field Theory}

Before we discuss the Affleck-Dine baryogenesis, let us review the
field theoretical methods for treating the quantum fluctuation and
back reaction of a dynamical quantum field.  The basic methods for
studying nonequilibrium phenomena were developed many years ago by 
Schwinger and Keldysh~\cite{schwinger}.  In nonequilibrium
situations we are interested in obtaining the equations of motion
for the expectation values and correlation functions of the quantum 
fields in an evolving quantum state or density matrix.  This is 
accomplished by tracking the time evolution of the density matrix, 
which corresponds to an initial valued problem for the Liouville 
equation.

In the Schr\"{o}dinger picture the evolution of the density matrix
$\rho$ is determined by
\begin{equation}
 \rho(t) = U(t, t_0)\, \rho(t_0)\, U^{-1}(t, t_0),
\end{equation}
where $U(t, t_0)$ is the time evolution operator:
\begin{equation}
 U (t, t_0) = {\cal T} \exp \left[ -i \int^t_{t_0} dt^{'} H (t')
 \right]\,
\end{equation}
for a time-dependent Hamiltonian $H$.  The symbol ${\cal T}$ means
to take the time-ordered product.  The expectation value of an
operator ${\cal O}$ in the Schr\"{o}dinger picture is given by
\begin{equation}
 \langle{\cal O}\rangle(t) = \frac{{\rm Tr}
 \left[ {U(t, t_0) \, \rho(t_0) \, U^{-1}(t, t_0) \, \cal O} \right]}
 {{\rm Tr} \left[ \rho(t_0) \right]}\, ,
\label{expectvalue1}
\end{equation}
and thus can be determined once the initial density matrix $\rho(t_0)$
is specified.

Consider the case in which the initial density matrix describes a
system in equilibrium. When a perturbation is switched on at time
$t_0$, the resulting time-dependent Hamiltonian can drive the
initial state out of equilibrium.  Just as in the usual perturbation
theory, we write $H = H_0 + H_{\rm int}$, where $H_0$ is the
unperturbed quadratic Hamiltonian and $H_ {\rm int}$ stands for the 
perturbations.  For technical convenience of treating this initial 
valued problem, we can model the dynamics by the following 
Hamiltonian:
\begin{equation}
 H(t)= \Theta (t_0 - t) H_0 (t_0) + \Theta (t - t_0) [H_0 (t) +
 H_{\rm int} (t)],
\label{t-depH}
\end{equation}
where $\Theta (t - t_0)$ is the Heaviside step function.  For
$ t <  t_0 $, the unperturbed Hamiltonian $ H_0 (t_0) $ is fixed
at time $t_0$ from which we can specify the initial thermal state
of the system, and for $ t > t_0$, the perturbation is switched on
in $ H_{\rm int} $.  Before the perturbation is switched on, the
system is assumed to be in equilibrium at a temperature $T = 1/\beta$
and is described by the initial density matrix
\begin{equation}
 \rho(t_0) = \exp[-\beta H_0 (t_0)] \, .
\label{densitymatrix-initial}
\end{equation}
The zero temperature limit for an initial vacuum state can be
obtained by taking $ T \rightarrow 0 $, as we shall consider later.

Notice that the initial density matrix can be expressed in terms of 
the time evolution operator along imaginary time in the distant past:
$\rho(t_0) = U (-\infty-i\beta, -\infty)$.  This allows us to write 
the expectation value in Eq.(\ref{expectvalue1}) as
\begin{equation}
 \langle{\cal O}\rangle(t) = \frac{{\rm Tr}
 \left[ U (-\infty-i \beta, -\infty) \, U^{-1}(t, -\infty) \, 
 {\cal O} \, U(t, -\infty) \right]}
 {{\rm Tr} \left[ U ( -\infty-i\beta, -\infty ) \right]},
\end{equation}
where we have inserted $U (t_0, -\infty) \, U^{-1} (t_0, -\infty) = 1$ 
to the right of $\rho(t_0)$ in the numerator, commuted $U (t_0, -\infty)$ 
with $\rho(t_0)$, and used the following composition property of the 
evolution operator: $U^{-1} (t_0, -\infty) \, U^{-1}(t, t_0) = [U(t, t_0) 
\, U (t_0, -\infty)]^{-1} = U^{-1}(t, -\infty)$.  Another insertion of 
$1 = U^{-1} (\infty, t) \, U (\infty, t)$ yields
\begin{equation}
 \langle{\cal O}\rangle(t) = \frac{{\rm Tr}
 \left[ U (-\infty-i \beta, -\infty) \, U^{-1} (\infty, -\infty) \, 
 U(\infty, t) \, {\cal O} \, U(t, -\infty) \right]}
 {{\rm Tr} \left[ U (-\infty-i\beta, -\infty) \right]} \, . 
\label{expectvalue}
\end{equation}
Although not apparent from the above expression, the time-dependent 
expectation value for ${\cal O}$ depends implicitly on $t_0$ through the 
perturbation $H_{\rm int}$ which is switched on at time $t_0$.  Because 
there are both forward and backward [given by $U^{-1} (\infty, -\infty)$] 
time evolutions, and because the initial density matrix is written as 
the time evolution operator along imaginary time, one is led to consider 
the following generating functional
\begin{equation}
 {\cal Z}[j^{+},j^{-}] = {\rm Tr} \left[ U(-\infty-i\beta, -\infty)
 \, U(-\infty, \infty, j^{-}) \, U(\infty, -\infty, j^{+}) \right]
\end{equation}
for deriving the nonequilibrium equations of motion for the
expectation values and correlation functions of the quantum fields.

In anticipation of the subsequent application to the Affleck-Dine
model~(\ref{model}) for baryogenesis, let us consider the case of
a scalar field $ \varphi $ as an example.  In this case the
generating functional above has the following path-integral
representation
\begin{eqnarray}
 && {\cal Z}[j^{+},j^{-}] = \int d\varphi d\varphi_1 d\varphi_2 \int
 {\cal D} \varphi^+ {\cal D} \varphi^- {\cal D} \varphi^{\beta} \,
 \exp \left( i \int^{\infty}_{-\infty}\, dt \, \int \, d^3 {\bf x} \,
 \left\{{\cal L} \left[ \varphi^+ \right] + j^+ \varphi^+ \right\}
 \right)
\nonumber \\
 && \times  \exp \left( -i \int^{\infty}_{-\infty}\, dt \, \int \,
 d^3 {\bf x} \, \left\{ {\cal L} \left[ \varphi^- \right] + j^-
 \varphi^- \right\} \right) \exp \left( i \int^{-\infty-i
 \beta}_{-\infty}\, dt \, \int \, d^3 {\bf x} \,\left\{ {\cal L}_0
 \left[ \varphi^{\beta} \right]
 \right\} \right) \,
\label{generatingfun}
\nonumber \\
\end{eqnarray}
with the  boundary conditions: $\varphi^+( {\bf x}, -\infty ) =
\varphi^{\beta} ( {\bf x}, -\infty-i\beta) = \varphi ({\bf x})$,
$\varphi^+( {\bf x}, \infty ) = \varphi^{-} ( {\bf x}, \infty) =
\varphi_1 ({\bf x})$, and $\varphi^-( {\bf x}, -\infty) =
\varphi^{\beta} ( {\bf x},- \infty) = \varphi_2 ({\bf x})$.  The
path integrals represented by the $\varphi^+$ and $\varphi^-$
fields correspond to the forward and backward time evolution,
respectively, while the one described by the $\varphi^{\beta}$
field corresponds to the time evolution along the path parallel to
the imaginary time axis.  The boundary conditions are in fact a
result of the trace as well as the bosonic nature of the operators.
Notice that the path integral corresponding to evolution in
imaginary time involves only the unperturbed quadratic Lagrangian
${\cal L}_0$ [see Eq.~(\ref{t-depH})], while the other path
integrals involve the full Lagrangian ${\cal L} = {\cal L}_0 +
{\cal L}_{\rm int}$, where ${\cal L}_{\rm int}$ stands for the
interaction Lagrangian that will be treated as a perturbation.

The introduction of sources in Eq.~(\ref{generatingfun}) allows
one to obtain expectation values involving the quantum field
operators by taking functional derivatives with respect to
the sources as follows:
\begin{equation}
 \varphi^+ \rightarrow - i \frac{\delta}{\delta j^+} \,\,\, ,
 \,\,\,\varphi^- \rightarrow i \frac{\delta}{\delta j^-} \,\,\, ,
\end{equation}
and then setting the sources to zero in the end. In fact, various
Green functions can be obtained by taking functional derivatives
of the generating function with respect to the appropriate
sources.  For instance, functional differentiation with respect
to the sources $j^+$ and $j^-$ can generate the time-ordered and
anti-time-ordered Green functions, respectively.

As for computing the real-time correlation functions of interest,
there is clearly no need to introduce any source along the path
parallel to the imaginary time axis since the Lagrangian along
this path involves only its unperturbed part ${\cal L}_0$.  In
the Schr\"{o}dinger picture, the ensemble average we implement
corresponds to an initial density matrix describing a state of
thermal equilibrium with respect to the unperturbed Hamiltonian.
Path integrals over ${\cal L}_0$ can be carried out exactly and we
can obtain the generating functional in terms of the nonequilibrium
propagators.  The temperature dependence enters through the
boundary conditions on the nonequilibrium propagators as we shall
see later.  Thus, the relevant generating functional for computing
real-time correlation functions can be obtained as
\begin{eqnarray}
 && {\cal Z}[j^{+},j^{-} ] =  \, \exp \left\{ i
 \int d^4x \left({\cal L}_{\rm int} \left[- i \frac{\delta}
 {\delta j^+} \right] - {\cal L}_{\rm int} \left[i
 \frac{\delta}{\delta j^-} \right] \right) \right\}  \,
 \nonumber \\
 && \,\,\,\,\,\,\,\,\,\, \times \exp \left\{ - \frac{1}{2}
 \int d^4x \, \int d^4x' \, \left[j^+ (x) \, G^{++}( x,x') \,
 j^+ (x') - j^+ (x)  \, G^{+-}( x,x')  \, j^- (x')
 \right.\right.
 \nonumber \\
 && \,\,\,\,\,\,\,\,\,\,\,\,\,\,\,\,\,\,\,\,\,\,\,\,\,\,\,\,\,\,
 \,\,\,\,\,\,\,\,\,\,\,\,\,\,\,\,\,\,\,\,\,\,\,\,\,\,\,\,\,\, \,\,
 \left.\left.  - j^- (x) \, G^{-+}( x,x') \, j^+ (x') + j^- (x) \,
 G^{--}( x,x') \, j^- (x') \right] \right\} \, ,
\label{neqgenfun}
\end{eqnarray}
where the nonequilibrium propagators corresponding to the forward
and backward time branches are given by 
\begin{eqnarray}
 G^{++} ({\bf x}, {\bf x'}; t,t') &=& \langle \varphi^+ ({\bf x},t)
 \varphi^+ ({\bf x'},t') \rangle
 \nonumber \\
 &=& G^{>} ({\bf x}, {\bf x'}; t,t') \, \Theta (t-t') \, +
 G^{<} ({\bf x}, {\bf x'}; t,t')\, \Theta (t'-t ) \, ,
 \nonumber \\
 G^{--} ({\bf x}, {\bf x'}; t,t') &=& \langle \varphi^- ({\bf x},t)
 \varphi^- ({\bf x'},t') \rangle
 \nonumber \\
 &=& G^{>} ({\bf x}, {\bf x'}; t,t') \, \Theta (t'-t) \, +
 G^{<} ({\bf x}, {\bf x'}; t,t') \, \Theta (t-t') \, ,
 \nonumber \\
 G^{+-} ({\bf x}, {\bf x'}; t,t') &=& \langle \varphi^+ ({\bf x},t)
 \varphi^- ({\bf x'},t') \rangle
 \nonumber \\
 &=& G^{<} ({\bf x}, {\bf x'}; t,t') = \langle \varphi ({\bf x'},t')
 \varphi ({\bf x},t) \rangle \, ,
 \nonumber \\
 G^{-+} ({\bf x}, {\bf x'}; t,t') &=& \langle \varphi^- ({\bf x},t)
 \varphi^+ ({\bf x'},t') \rangle
 \nonumber \\
 &=& G^{>} ({\bf x}, {\bf x'}; t,t') = \langle \varphi ({\bf x},t)
 \varphi ({\bf x'},t') \rangle \, .
\label{neqpropagator}
\end{eqnarray}
Since time translational invariance is lost for a system that is
out of equilibrium, the Green functions defined above will in
general depend on the time $t$ and $t'$ separately.

The functions $G^{>}$ and $G^{<}$ can be obtained from computing
the 2-point correlation functions of the quantum field $\varphi$,
which can be expressed in terms of the usual creation and
annihilation operators with mode functions that are solutions to
the homogeneous equations of motion for the quadratic Lagrangian
${\cal L}_0$.  The explicit derivation for them will be shown in
the next section.  As a result of the assumed initial thermal
state mentioned above, they obey the boundary condition
\begin{equation}
 G^{<} ({\bf x}, {\bf x'}; -\infty, t') = G^{>} ({\bf x},
 {\bf x'}; -\infty-i\beta, t') \, ,
\end{equation}
which corresponds to the Kubo-Martin-Schwinger (KMS) condition
at a large negative time before the pertubation is switched
on~\cite{bellac}.

In summary, the generating functional~(\ref{neqgenfun}) can be
used to obtain the Feynman rules that define the perturbative
expansion for weak couplings.  There are four sets of
nonequilibrium propagators in the forms given in
Eq.~(\ref{neqpropagator}).  Two sets of interaction vertices
defined by the interaction Lagrangian ${\cal L}_{\rm int}$
involve fields in the forward branch and those in the backward
branch.  Notice that there exists a relative minus sign difference
between these two types of vertices.  The calculation of the
combinatorial factors is the same as in the usual quantum field
theory.  This constitutes the basic tool for our studies.

\section{Complex Scalar Field in an Expanding Space-time}

Since the interaction rate for scattering processes between the
scalar fermion fields and the thermal bath in the early Universe
is found~\cite{afdine} to be of the same order of magnitude as
their decay rate $\Gamma_\Phi^{-1}$ (see Sec.~II), the scalar
fermions are essentially decoupled from the bath for a time scale
much longer than the oscillation period of the scalar fields:
$m_{\Phi}^{-1} \ll t \ll \Gamma_\Phi^{-1}$.  We can therefore
apply the method of nonequilibrium quantum field theory with zero
temperature to study Affleck-Dine baryogenesis by considering
the model of Eq.~(\ref{model}).

The relevant action can  be written as
\begin{equation}
 {\cal S} = \int d^4 x \, \sqrt{-g} \, \left[ \, g_{\mu \nu}
 \left(\partial^{\mu} \Phi^{\dagger} \right) \left(\partial^{\nu}
 \Phi \right) - m^2_{\Phi} \Phi^{\dagger} \Phi - i \lambda
 \left(\Phi^4 - \Phi^{\dagger \, 4} \right) \, \right] \, .
\end{equation}
The background geometry is governed by the spatially flat
Robertson-Walker metric,
\begin{equation}
 d^2 s = d^2 t - a^2 (t) d^2 {\bf x} \,,
\end{equation}
where $a(t)$ is a scale factor.  It will be convenient to
introduce the conformal time $\eta$ defined as $ d \eta = a^{-1}d t$
and the conformally rescaled field, $\chi = a \, \Phi$, whereby
the above action can be written as that of a complex scalar field
in flat space-time with a time-dependent mass term.  To see this, we
express the complex scalar field $\chi$ in terms of its real and
imaginary parts: $\chi = (1/\sqrt{2}) \, (\chi_1 + i \chi_2)$.  The
action becomes
\begin{eqnarray}
 {\cal S} & = & \int d\eta \, d^3{\bf x} \, {\cal L} \left[
 \chi_1,\chi_2 \right] \nonumber \\
 &=&  \int d\eta \, d^3{\bf x} \, \left[\frac{1}{2} \chi_1^{' \, 2}
 - \frac{1}{2} \left( {\vec \bigtriangledown} \chi_1 \right)^2 +
 \frac{1}{2} \chi_2^{' \, 2} - \frac{1}{2} \left({\vec
 \bigtriangledown} \chi_2 \right)^2 - U(\chi_1,\chi_2) \right] \, ,
\end{eqnarray}
where $\chi_i' \equiv \partial \chi_i/{\partial \eta}$ and the new
potential
\begin{equation}
 U(\chi_1,\chi_2) = \frac{1}{2} m^2_{\chi} (\eta) \left(
 \chi_1^2+\chi_2^2 \right) - 2 \lambda \chi_1 \chi_2 \left(
 \chi_1^2-\chi_2^2 \right)
\end{equation}
has the time-dependent mass term with $m^2_{\chi} (\eta) = m^2_{\Phi}
a^2 - (a^{''}/a)$.

Since we are interested in the effects of the quantum fluctuations
of the scalar fields $\chi_1$ and $\chi_2$, it will be useful to
express each of them in terms of its expectation value with respect
to an initial density matrix that will be specified later and the
fluctuation field around this average value:
\begin{eqnarray}
 \chi_1^{\pm} ( {\bf x},\eta) &=&\chi_1^0 (\eta) + \tilde
 {\chi}_1^{\pm} ( {\bf x},\eta ) \, , \,\,\,\,\,  \langle
 \chi_1^{\pm} ( {\bf x},\eta)
 \rangle= \chi_1^0 (\eta) \, ; \nonumber \\
 \chi_2^{\pm} ( {\bf x},\eta) &=& \chi_2^0 (\eta) + \tilde
 {\chi}_2^{\pm} ( {\bf x},\eta ) \, , \,\,\,\,\,  \langle
 \chi_2^{\pm} ( {\bf x},\eta) \rangle= \chi_2^0 (\eta) \, .
\end{eqnarray}
We have listed the fields belonging to the forward $(+)$ and
backward $(-)$ time branches explicitly.  The reason for shifting
the ($\pm$) fields by the same configuration is that the field
expectation values enter in the time evolution operator as
c-number background fields, and the evolution forward and backward
are considered in this background~\cite{boyan1}.  Note that
$\chi_1^0$ and $\chi_2^0$ are time-dependent for the
nonequilibrium problem. We will implement later the corresponding
tadpole conditions, $\langle \tilde{\chi}_1^{\pm} ( {\bf x},\eta)
\rangle=0 $  and $ \langle \tilde{\chi}_2^{\pm} ( {\bf x},\eta)
\rangle =0 $, to derive the equations of motion for these field
expectation values. Expanding the action in terms of the
fluctuation fields $\tilde{\chi}_1 ({\bf x},\eta)$ and
$\tilde{\chi}_2 ({\bf x},\eta)$, the Lagrangian density becomes
\begin{widetext}
\begin{eqnarray}
 {\cal L} \left[\chi_1^0+\tilde{\chi}_1^+,\chi_2^0+\tilde{\chi}_2^+
 \right] &=& {\cal L} \left[\chi_1^0,\chi_2^0 \right] +
 {\cal L} \left[\tilde{\chi}_1^+,\tilde{\chi}_2^+ \right]
 \nonumber \\
 &-& \tilde{\chi}_1^+ \left\{ \chi_1^{0 \, ''} + m^2_{\chi} (\eta)
 \chi_1^0 - 2 \lambda \left[ 3 \left(\chi_1^0 \right)^2 \chi_2^0
 - \left(\chi_2^0 \right)^3 \right] \right\}
 \nonumber \\
 &-& \tilde{\chi}_2^+ \left\{ \chi_2^{0 \, ''} + m^2_{\chi} (\eta)
 \chi_2^0 + 2 \lambda \left[ 3 \chi_1^0 \left(\chi_2^0 \right)^2
 - \left(\chi_1^0 \right)^3 \right] \right\}
 \nonumber \\
 &+& 6 \lambda \chi_1^0 \chi_2^0 \left(\tilde{\chi}_1^{+  \, 2} -
 \tilde{\chi}_2^{+  \, 2} \right) + 6 \lambda \tilde{\chi}_1^+
 \tilde{\chi}_2^+ \left[ \left(\chi_1^0 \right)^2 - \left(\chi_2^0
 \right)^2 \right]
 \nonumber \\
 &+& 2 \lambda \left(\chi_2^0 \tilde{\chi}_1^{+  \, 3} - \chi_1^0
 \tilde{\chi}_2^{+  \, 3} + 3 \chi_1^0 \tilde{\chi}_1^{+  \, 2}
 \tilde{\chi}_2^+ - 3 \chi_2^0 \tilde{\chi}_1^+
 \tilde{\chi}_2^{+  \, 2} \right)
\end{eqnarray}
\end{widetext}
and similarly for ${\cal L} \left[\chi_1^0+\tilde{\chi}_1^-,\chi_2^0
+\tilde{\chi}_2^- \right]$.

We now introduce a Hartree factorization to take account of the
quantum fluctuations.  The Hartree factorization can be employed
for both the $(+)$ and $(-)$ fields  as follows:
\begin{eqnarray}
 && \tilde{\chi}_1^3 \approx 3 \langle \tilde{\chi}_1^2 \rangle
 \tilde{\chi}_1 \, , \,\,\, \,\,\, \,\,\, \,\,\,\,\,\, \,\,\,\,\,\,
 \,\,\, \,\,\, \,\,\, \,\,\, \,\,\, \,\,\,\,\,\,\,\,\,\,\,\,\,\,\,
 \tilde{\chi}_2^3 \approx 3 \langle
 \tilde{\chi}_2^2 \rangle \tilde{\chi}_2 \, , \nonumber \\
 && \tilde{\chi}_1^2 \tilde{\chi}_2 \approx 2 \langle
 \tilde{\chi}_1 \tilde{\chi}_2 \rangle \tilde{\chi}_1 + \langle
 \tilde{\chi}_1^2 \rangle \tilde{\chi}_2 \, ,
 \,\,\,\,\,\,\,\,\,\,\,\,\,\,\, \tilde{\chi}_1
 \tilde{\chi}_2^2\approx 2 \langle \tilde{\chi}_1
 \tilde{\chi}_2 \rangle \tilde{\chi}_2 + \langle
 \tilde{\chi}_2^2 \rangle \tilde{\chi}_1 \, , \nonumber \\
 && \tilde{\chi}_1^3 \tilde{\chi}_2  \approx 3 \langle
 \tilde{\chi}_1 \tilde{\chi}_2 \rangle \tilde{\chi}_1^2 + 3 \langle
 \tilde{\chi}_1^2 \rangle   \tilde{\chi}_1 \tilde{\chi}_2 \, ,
 \,\,\,\,\,\,\, \tilde{\chi}_1 \tilde{\chi}_2^3 \approx 3
 \langle \tilde{\chi}_1 \tilde{\chi}_2 \rangle \tilde{\chi}_2^2 + 3
 \langle \tilde{\chi}_2^2 \rangle    \tilde{\chi}_1 \tilde{\chi}_2
 \, ,
\end{eqnarray}
where the one-loop corrections to the 2-point Green functions
of $\tilde{\chi}_1$ and $\tilde{\chi}_2$ are cancelled by the
counterterms introduced above.  We would like to point out that
the Hartree approximation we adopt here is uncontrolled in the
model of Eq.~(\ref{model}).  It will be controlled, for example,
in alternative models in which the scalar fields are in the vector
representation of the symmetry group $O(N)$ as the Hartree
approximation is then equivalent to the large-$N$ approximation.
Our justification for using this approximation here is simply
based on the virtue that it provides a nonperturbative framework
to treat the quantum fluctuations by solving the evolution
equations self-consistently.

After applying the above factorizations, the nonequilibrium
Lagrangian density can be written as
\begin{eqnarray}
 && {\cal L} \left[\chi_1^0+\tilde{\chi}_1^+,\chi_2^0+\tilde{\chi}_2^+ 
 \right] - {\cal L} \left[\chi_1^0+\tilde{\chi}_1^-,\chi_2^0 +
 \tilde{\chi}_2^- \right]
\nonumber \\
 &=& {\cal L}_{0} \left[\tilde{\chi}_1^+, \tilde{\chi}_2^+ \right] -
 {\cal L}_{0} \left[\tilde{\chi}_1^-, \tilde{\chi}_2^- \right] +
 {\cal L}_{\rm int} \left[\tilde{\chi}_1^+, \tilde{\chi}_2^+ \right] -
 {\cal L}_{\rm int} \left[\tilde{\chi}_1^-, \tilde{\chi}_2^- \right]
 \, ,
\label{HartreeL}
\end{eqnarray}
where
\begin{eqnarray}
 {\cal L}_{0} \left[\tilde{\chi}_1^+,\tilde{\chi}_2^+ \right]
 &=& \frac{1}{2} \tilde{\chi}_1^{+ ' \, 2} - \frac{1}{2} \left({\vec
 \bigtriangledown} \tilde{\chi}_1^+ \right)^2 - \frac{1}{2}
 {\cal M}_{\chi_1}^2 (\eta) \tilde{\chi}_1^{+  \, 2} \,
 \nonumber \\
 ~~&+& \frac{1}{2} \tilde{\chi}_2^{+ ' \, 2} - \frac{1}{2} \left(
 {\vec \bigtriangledown} \tilde{\chi}_2^+ \right)^2 - \frac{1}{2}
 {\cal M}_{\chi_2}^2 (\eta) \tilde{\chi}_2^{+ \, 2} \, ,
\label{HartreeL0}
\end{eqnarray}
with
\begin{eqnarray}
 {\cal M}_{\chi_1}^2 (\eta) &=& m^2_{\chi} (\eta) - 12 \lambda
 \chi^0_1 \chi^0_2 - 12 \lambda \langle \tilde{\chi}_1 \tilde{\chi}_2
 \rangle \, ,
 \nonumber \\
 {\cal M}_{\chi_2}^2 (\eta) &=& m^2_{\chi} (\eta) + 12 \lambda
 \chi^0_1 \chi^0_2 + 12 \lambda \langle \tilde{\chi}_1 \tilde{\chi}_2
 \rangle \, ,
\label{Hartreemass}
\end{eqnarray}
and
\begin{equation}
 {\cal L}_{\rm int} \left[\tilde{\chi}_1^+,\tilde{\chi}_2^+ \right]
 = - \alpha_1 (\eta) \tilde{\chi}_1^+ - \alpha_2 (\eta)
 \tilde{\chi}_2^+ + \alpha_{12} (\eta) \tilde{\chi}_1^+
 \tilde{\chi}_2^+ \, ,
\label{HartreeLint}
\end{equation}
with the couplings
\begin{eqnarray}
 \alpha_1 (\eta) &=& \chi_1^{0 \, ''} + m^2_{\chi} (\eta) \chi_1^0
 - 2 \lambda \left[3 \left( \chi_1^0 \right)^2 \chi_2^0 - \left(
 \chi_2^0 \right)^3 \right] - 6 \lambda \chi_2^0  \left(\langle
 \tilde{\chi}_1^2 \rangle - \langle \tilde{\chi}_2^2 \rangle \right)
 - 12 \lambda \chi_1^0 \langle \tilde{\chi}_1 \tilde{\chi}_2 \rangle
 \, ,
 \nonumber \\
 \alpha_2 (\eta) &=& \chi_2^{0 \, ''} + m^2_{\chi} (\eta) \chi_2^0
 + 2 \lambda \left[3 \chi_1^0 \left( \chi_2^0 \right)^2 - \left(
 \chi_1^0 \right)^3 \right] - 6 \lambda \chi_1^0 \left(\langle
 \tilde{\chi}_1^2 \rangle - \langle \tilde{\chi}_2^2 \rangle \right)
 + 12 \lambda \chi_2^0 \langle \tilde{\chi}_1 \tilde{\chi}_2 \rangle
 \, ,
\nonumber \\
 \alpha_{12} (\eta) &=& 6 \lambda \left[ \left( \chi^0_1 \right)^2
 - \left( \chi^0_2 \right)^2 \right] + 6 \lambda \left( \langle
 \tilde{\chi}_1^2 \rangle - \langle \tilde{\chi}_2^2 \rangle \right)
 \, .
\label{Hartreepot}
\end{eqnarray}
We have used the fact that correlation functions of fields
evaluated at the same space-time point in the forward and
backward time branches are equal.  This can be seen from
Eq.~(\ref{neqpropagator}) by identifying $({\bf x}, t)$
with $({\bf x}', t')$.  In addition, due to the spatial
translational invariance, the correlation functions depend
only on time and should be understood as:
$\langle \tilde{\chi}_{i}^{+ \, 2} ({\bf x}, \eta) \rangle  =
\langle \tilde{\chi}_{i}^{- \, 2} ({\bf x}, \eta) \rangle =
\langle \tilde{\chi}_{i}^{2} \rangle (\eta) \ (i=1,2)$, $\langle
\tilde{\chi}_{1}^{+} ({\bf x}, \eta) \tilde{\chi}_{2}^{+} ({\bf
x}, \eta) \rangle = \langle \tilde{\chi}_{1}^{-} ({\bf x}, \eta)
\tilde{\chi}_{2}^{-} ({\bf x}, \eta) \rangle = \langle
\tilde{\chi}_{1} \tilde{\chi}_{2} \rangle (\eta)$.

In Eq.~(\ref{HartreeL}), ${\cal L}_{0}$ is the unperturbed
quadratic Lagrangian with respect to which the nonequilibrium
Green functions will be defined and the interaction Lagrangian
${\cal L}_{\rm int}$ contains the linear terms in
$\tilde{\chi}_{i}$ as well as the $\tilde{\chi}_1 \tilde{\chi}_2$
terms which will be treated as perturbations.  The presence of the
nondiagonal $\tilde{\chi}_1 \tilde{\chi}_2$ terms implies that
these fields do not form a good basis to construct the
nonequilibrium propagators in a perturbative expansion.  One may
therefore consider invoking the Bogoliubov transformation which is
a canonical transformation in the sense that it leaves the measure
of the path integral invariant.  In this case, the transformation
is a two-dimensional field rotation between the old and new basis
fields with one parameter specified by the rotation angle.  The
angle for the field rotation can be determined so as to rid the
nondiagonal term in the potential energy of the Lagrangian.  In
the presence of the time-dependent background fields, it turns out
that the Bogoliubov transformation acquires a time-dependent
rotation angle, which in turn gives rise to a new kind of unwanted
nondiagonal terms in the kinetic energy.  Thus, it seems that the
Bogoliubov transformation fails to diagonalize the Lagrangian in
the presence of the time-dependent background fields.  To proceed,
we can further approximate the term $\tilde{\chi}_1
\tilde{\chi}_2$ by its mean value $\langle \tilde{\chi}_1
\tilde{\chi}_2 \rangle$.  In other words, we write
\begin{equation}
 \tilde{\chi}_1 \tilde{\chi}_2 = \langle \tilde{\chi}_1 \tilde{
 \chi}_2 \rangle + \left(\tilde{\chi}_1 \tilde{\chi}_2 -\langle
 \tilde{\chi}_1 \tilde{\chi}_2 \rangle \right) \, ,
\end{equation}
and treat the terms in the parenthesis as a perturbation.

The fluctuation fields can be decomposed in terms of their Fourier
modes with time-dependent mode functions as
\begin{eqnarray}
 \tilde{\chi}_i ({\bf x},\eta) &=& \frac{1}{\sqrt{\Omega}} \,
 \sum_{\bf k} \, \tilde{\chi}_{i , {\bf k}} (\eta) \, e^{i {\bf k}
 \cdot {\bf x}}
 \nonumber \\
 &=& \frac{1}{\sqrt{\Omega}} \, \sum_{\bf k} \, \left[ \,
 a_{i , {\bf k}} \, f_{i , {\bf k}}(\eta) +
 \, a^{\dagger}_{i , {\bf -k}} \, f^{*}_{i , {\bf -k}} (\eta) \,
 \right]\,  e^{i {\bf k} \cdot {\bf x}} \, , \,\,\,\,\, i=1,2 \, ,
\label{modeexpand}
\end{eqnarray}
where $a_{i , {\bf k}}$ and $a^{\dagger}_{i , {\bf -k}}$ are,
respectively, the annihilation and creation operators for the
fluctuation field $\tilde{\chi}_i$ with momentum ${\bf k}$.  The
equations of motion for the mode functions $f_{i , {\bf k}}$ can
be read off from  the quadratic Lagrangian ${\cal L}_0$ in
Eq.~(\ref{HartreeL0}) as
\begin{equation}
 \left[ \frac{d^2}{d \eta^2 } + k^2 + {\cal M}_{\chi_i}^2 (\eta)
 \right] f_{i , {\bf k}} (\eta) = 0 \, , \,\,\,\,\, i=1,2 \, ,
\label{modeeq}
\end{equation}
where $k^2 = {\bf k} \cdot {\bf k}$.  The nonequilibrium propagators
can also be expressed in terms of the mode functions:
\begin{eqnarray}
 G^{++}_{i , {\bf k}} (\eta , \eta' ) &=& G^{>}_{i , {\bf k}}
 (\eta , \eta') \, \Theta (\eta - \eta') \, + G^{<}_{i , {\bf k}}
 (\eta , \eta' )\, \Theta (\eta' -\eta )\, , \nonumber \\
 G^{--}_{i , {\bf k}} (\eta  , \eta' ) &=& G^{>}_{i , {\bf k}}
 (\eta , \eta' ) \, \Theta (\eta' - \eta) \, + G^{<}_{i , {\bf k}}
 (\eta , \eta' )\,  \Theta (\eta - \eta' ) \, ,\nonumber \\
 G^{+-}_{i , {\bf k}} (\eta , \eta' ) &=& G^{<}_{i , {\bf k}}
 (\eta , \eta' ) \, , \nonumber \\
 G^{-+}_{i , {\bf k}} (\eta , \eta' ) &=& G^{>}_{i , {\bf k}}
 (\eta , \eta' ) \, ,  \nonumber \\
 G^{>}_{i , {\bf k}} (\eta , \eta' ) &=& \int d^3 {\bf x} \,
 \langle \tilde{\chi}_{i} ({\bf x},\eta) \tilde{\chi}_{i} ({\bf
 0} ,\eta') \rangle \, e^{-i {\bf k} \cdot  {\bf x} } \nonumber \\
 &=& f_{i , {\bf k}} (\eta) f^{*}_{i , {\bf k}} (\eta' ) \, ,
 \nonumber \\
 G^{<}_{i , {\bf k}} (\eta , \eta^{'} ) &=& \int d^3 {\bf x} \,
 \langle \tilde{\chi}_{i} ({\bf 0},\eta') \tilde{\chi}_{i} ({\bf
 x},\eta) \rangle \, e^{-i {\bf k} \cdot  {\bf x} } \nonumber \\
 &=& f_{i , {\bf k}} (\eta') f^{*}_{i , {\bf k}} (\eta ) \, ,
\label{neqprops}
\end{eqnarray}
where the correlations of the fluctuation fields are taken with
respect to a vacuum state that is annihilated by the $a_{i , {\bf
k}}$ of Eq.~(\ref{modeexpand}).

The equations of motion for the expectation values of the fields
can be obtained from the tadpole conditions. From the condition
$\langle \tilde{\chi}^{+}_{1} ( {\bf x}, \eta)\rangle=0$, the
lowest-order contributions to the equation of motion are obtained
from the terms linear in $\tilde{\chi}_{1}^{+}$ in the interaction
Lagrangian~(\ref{HartreeLint}).  We find
\begin{equation}
 \int d\eta \left[ - \alpha_1 (\eta) \langle \tilde{\chi}_{1}^{+}
 \tilde{\chi}_{1}^{+} \rangle + \alpha_1 (\eta) \langle
 \tilde{\chi}_{1}^{+} \tilde{\chi}_{1}^{-} \rangle \right] = 0 \, ,
\end{equation}
where the spatial arguments have been suppressed.  Since the
correlation functions $\langle \tilde{\chi}_{1}^{+}
\tilde{\chi}_{1}^{+} \rangle$ and $\langle \tilde{\chi}_{1}^{+}
\tilde{\chi}_{1}^{-} \rangle$ are linearly independent, the above
equation leads to the equation of motion for $\chi_1^0$:
\begin{equation}
 \chi_1^{0 \, ''} + m^2_{\chi} (\eta) \chi_1^0 - 2 \lambda \left[
 3 \left( \chi_1^0 \right)^2 \chi_2^0 - \left( \chi_2^0 \right)^3
 \right] - 6 \lambda \chi_2^0 \left( \langle \tilde{\chi}_1^2 \rangle
 - \langle \tilde{\chi}_2^2 \rangle \right) - 12 \lambda \chi_1^0
 \langle \tilde{\chi}_1 \tilde{\chi}_2 \rangle = 0 \, ,
\label{chi1eq}
\end{equation}
where Eq.~(\ref{Hartreepot}) has been used.  It is straightforward
to see that implementing the condition $\langle \tilde{\chi}^{-}_{1}
({\bf x}, \eta)\rangle = 0$ will lead to the same equation of
motion for $\chi_1^0$.  Following the similar procedure with the
conditions $\langle \tilde{\chi}_{2}^{\pm} ({\bf x}, \eta) \rangle
= 0$, we find the equation of motion for $\chi_2^{0}$ to be
\begin{equation}
 \chi_2^{0 \, ''} + m^2_{\chi} (\eta) \chi_2^0 + 2 \lambda \left[
 3 \chi_1^{0} \left( \chi_2^0 \right)^2 - \left( \chi_1^0 \right)^3
 \right] - 6 \lambda \chi_1^0 \left( \langle \tilde{\chi}_1^2 \rangle
 - \langle \tilde{\chi}_2^2 \rangle \right) + 12 \lambda \chi_2^0
 \langle \tilde{\chi}_1 \tilde{\chi}_2 \rangle = 0 \, .
\label{chi2eq}
\end{equation}

The quantum fluctuations $\langle \tilde{\chi}^2_1 \rangle$ and
$\langle \tilde{\chi}^2_2 \rangle$ can be determined
self-consistently and have the form
\begin{equation}
 \langle \tilde{\chi}^2_i \rangle (\eta) = \int \, \frac{d^3 {\bf
 k}}{ ( 2 \pi)^3 } \, \mid f_{i , {\bf k}} (\eta) \mid^2 ,
\label{chi-chifluc}
\end{equation}
where the infinite volume limit has been taken.  On the other
hand, the term $\langle \tilde{\chi}_1 \tilde{\chi}_2 \rangle$
vanishes at tree level and one has to include an interaction
vertex from the perturbations to obtain its loop corrections.  To
lowest order in perturbation theory, the vertex $6 i \lambda
\left[ \left( \chi^0_1 \right)^2 - \left( \chi^0_2 \right)^2
\right] \tilde{\chi}_1 \tilde{\chi}_2$ is included and one finds
that
\begin{eqnarray}
 \langle \tilde{\chi}_1 \tilde{\chi}_2 \rangle (\eta) &=& \int
 \frac{d^3 {\bf k}}{(2 \pi)^3} \, \langle \tilde{\chi}_{1 , {\bf k}}
 (\eta) \tilde{\chi}_{2 , {\bf -k}} (\eta) \rangle \,
 \nonumber \\
 &=& 6 i \, \lambda \int_{\eta_0}^{\eta} d\eta^{'} \left\{ \left[
 \chi_1^0 (\eta') \right]^2 - \left[ \chi_2^0 (\eta') \right]^2
 \right\} \left[ \langle \tilde{\chi}^{+}_{1 , {\bf k}} (\eta)
 \tilde{\chi}^{+}_{1 , {\bf -k}} (\eta') \rangle
 \langle \tilde{\chi}^{+}_{2 , {\bf -k}} (\eta) \tilde{\chi}^{+}_
 {2 , {\bf k}} (\eta') \rangle \right.
 \nonumber \\
 && \left.   \,\,\,\,\,\,\,\,\,\,\,\,\,\,\,\,\,\,\,\,\,\,\,\,\,\,
 \,\,\,\,\,\,\,\,\,\,\,\,\,\,\,\,\,\,\,\,\,\,\,\,\,\,\,\,\,\,\,\,\,\,
 - \langle \tilde{\chi}^{+}_{1 , {\bf k}} (\eta)
 \tilde{\chi}^{-}_{1 , {\bf -k}} (\eta') \rangle
 \langle \tilde{\chi}^{+}_{2 , {\bf -k}} (\eta) \tilde{\chi}^{-}_
 {2 , {\bf k}} (\eta') \rangle \right] \,
 \nonumber \\
 &=& 6 i \, \lambda \int_{\eta_0}^{\eta} d\eta^{'} \left\{ \left[
 \chi_1^0 (\eta') \right]^2 - \left[ \chi_2^0 (\eta') \right]^2
 \right\} \left[ G^{>}_{1 , {\bf k}} (\eta , \eta') \,
 G^{>}_{2 , {\bf k}} (\eta , \eta') \right.
 \nonumber \\
 && \left. \,\,\,\,\,\,\,\,\,\,\,\,\,\,\,\,\,\,\,\,\,\,\,\,\,\,
 \,\,\,\,\,\,\,\,\,\,\,\,\,\,\,\,\,\,\,\,\,\,\,\,\,\,\,\,\,\,\,\,\,\,
 - G^{<}_{1 , {\bf k}} (\eta , \eta') \, G^{<}_{2 , {\bf k}}
 (\eta , \eta') \right] \, \Theta (\eta - \eta') \, .
\end{eqnarray}
As mentioned earlier, the $\eta_0$ dependence arises from the
perturbations being switched on at time $\eta_0$.  We have shown
above only the calculation (second equality) for the $\tilde{\chi}_1^+
\tilde{\chi}_2^+$ correlation.  However, as we have stated after
Eq.~(\ref{Hartreepot}), the final result is also applicable to
$\langle \tilde{\chi}_{1}^{-} \tilde{\chi}_{2}^{-} \rangle$.  In
terms of the mode functions, this correlation function becomes
\begin{equation}
 \langle \tilde{\chi}_1 \tilde{\chi}_2 \rangle (\eta) = 6 \, i \lambda
 \, \int_{\eta_0}^{\eta} \, d\eta' \, \left\{ \left[ \chi_1^0 (\eta')
 \right]^2 - \left[ \chi_2^0 (\eta') \right]^2 \right\} \left[
 f_{1 , {\bf k}} (\eta) f^{*}_{1 , {\bf k}} (\eta')
 f_{2 , {\bf k}} (\eta) f^{*}_{2 , {\bf k}} (\eta') - {\rm c. c.}
 \right] \, .
\label{chi1-chi2fluc}
\end{equation}
Equations (\ref{modeeq}), (\ref{chi1eq}), (\ref{chi2eq}),
(\ref{chi-chifluc}) and (\ref{chi1-chi2fluc}) form the full set of
coupled equations from which we will solve for the field
expectation values and the mode functions.

\section{Nonequilibrium Effects and Baryon Asymmetry}

The proper normalization to measure baryon asymmetry is given by
the particle number density.  Let us consider the initial state 
at time $t_0$ to be specified by the adiabatic modes for uncoupled
harmonic oscillations defined in cosmic time under Hartree
approximations.  The corresponding initial particle number density
can be expressed as
\begin{eqnarray}
 \hat{n} (t_0) &=& \int \frac{d^3 {\bf k}}{ ( 2\pi)^3} \left\{ \frac
 {a^3(t_0)}{2\, {\cal W}_{1, {\bf k}} (t_0)} \left[ \dot{\Phi}_{1 ,
 {\bf k}} (t_0) \dot{\Phi}_{1 , {\bf -k}}(t_0)  +{\cal W}^2_{1,
 {\bf k}} (t_0) \, {\Phi}_{1 , {\bf k}} (t_0) {\Phi}_{1 , {\bf -k}}
 (t_0)\right] \right. \nonumber \\
 && \left. + \frac{ a^3
 (t_0)}{2 \, {\cal W}_{2, {\bf k}} (t_0)} \left[ \dot{\Phi}_{2 ,
 {\bf k}} (t_0) \dot{\Phi}_{2 , {\bf -k}} (t_0) +{\cal W}^2_{2,
 {\bf k}} (t_0) \, {\Phi}_{2 , {\bf k}} (t_0) {\Phi}_{2 , {\bf -k}}
 (t_0)\right]-1 \right\} \, ,
\end{eqnarray}
where $\Phi = (\Phi_1 +i \Phi_2)/\sqrt{2}$.
The instantaneous frequencies for the real and imaginary part of the
complex scalar field $\Phi$ can be gleaned from Eqs.~(\ref{modeeq})
and (\ref{Hartreemass}) to be
\begin{eqnarray}
 {\cal W}_{1, {\bf k}} (t_0) &=& \left[ \frac{k^2}{a^2 (t_0) } +
 m^2_{\Phi} - 12\lambda \, a^{-2} (t_0) \left(\chi_1^0 (t_0) \chi^0_2
 (t_0) + \langle \tilde{\chi}_1 \tilde{\chi}_2 \rangle (t_0) \right)
 \right]^{\frac{1}{2}} \, ,
 \nonumber \\
 {\cal W}_{2, {\bf k}} (t_0) &=& \left[ \frac{k^2}{a^2 (t_0) } +
 m^2_{\Phi} + 12\lambda \, a^{-2} (t_0) \left(\chi_1^0 (t_0) \chi^0_2
 (t_0) + \langle \tilde{\chi}_1 \tilde{\chi}_2 \rangle (t_0) \right)
 \right]^{\frac{1}{2}} \, .
\end{eqnarray}
The expectation value of the number operator with respect to an
initial vacuum state as defined after Eq.~(\ref{neqprops}) evolves
in time and has the following form in the Heisenberg picture:
\begin{eqnarray}
\langle \hat{n} \rangle (t) &=& \int \frac{d^3 {\bf k}}{ ( 2\pi)^3}
 \left\{ \frac{ a^3
 (t_0)}{{ 2 \cal W}_{1, {\bf k}} (t_0)} \left[ \langle \dot{\Phi}_{1 ,
 {\bf k}} (t) \dot{\Phi}_{1 , {\bf -k}}(t)  \rangle +{\cal W}^2_{1,
 {\bf k}} (t_0) \,  \langle {\Phi}_{1 , {\bf k}} (t) {\Phi}_{1 ,
 {\bf -k}} (t) \rangle \right] \right. \nonumber \\
 && \left. + \frac{ a^3 (t_0)}{2 {\cal W}_{2, {\bf k}} (t_0)} \left[
 \langle \dot{\Phi}_{2 , {\bf k}} (t) \dot{\Phi}_{2 , {\bf -k}} (t)
 \rangle +{\cal W}^2_{2, {\bf k}} (t_0) \, \langle {\Phi}_{2 , {\bf k}}
 (t) {\Phi}_{2 , {\bf -k}} (t) \rangle \right]-1 \right\} \, .
 \end{eqnarray}
In terms of the conformally rescaled fields, the expectation value
of the particle number density can be written as the mean value
constructed from the expectation values of the quantum fields plus
its quantum fluctuations:
\begin{eqnarray}
 \langle \hat{n} \rangle (\eta) &\equiv& n(\eta) = n_{0} (\eta) +
 n_{\rm q} (\eta) \, ,
 \nonumber \\
 n_{ 0} (\eta) &=&  \frac{ a^3 (\eta_0)} {2 \, {\cal W}_{1, {\bf 0}}
 (\eta_0)\, a^4 (\eta)}  \, \left\{ \left[\frac{d}{d\eta} \chi_1^0 (\eta)
 \right]^2 - \frac{a'(\eta)}{a (\eta)} \, \frac{d}{d\eta} \left[\chi_1^{0}
 (\eta) \right]^2 + \Omega_{1 , {\bf k}} (\eta) \, \left[\chi_1^{0} (\eta)
 \right]^2 \right\}
 \nonumber\\
 && + \left( \chi_1 \rightarrow \chi_2 \, ; \, \Omega_{1 , {\bf k}}
 (\eta) \rightarrow \Omega_{2 , {\bf k}} (\eta) \right) \, ,
 \nonumber \\
 n_{\rm q} (\eta) &=&  \int \,   \frac{d^3 {\bf k}}{ ( 2\pi)^3} \,
 \left\{ \frac{a^3 (\eta_0) }{ 2 \, {\cal W}_{1, {\bf k}} (\eta_0)
 \, a^4 (\eta)} \, \left[ \langle \tilde{\chi}'_{1 , {\bf k}}
 (\eta) \tilde{\chi}'_{1 , {\bf -k}} (\eta) \rangle \, - \frac{a'
 (\eta)}{ a(\eta)} \, \left( \frac{d}{d\eta} \langle
 \tilde{\chi}_{1 , {\bf k}} (\eta) \tilde{\chi}_{1 , {\bf -k}}
 (\eta) \rangle \right) \right. \right.
 \nonumber \\
 && \left. \left. + \Omega_{1 , {\bf k}} (\eta) \, \langle
 \tilde{\chi}_{1 , {\bf k}} (\eta) \tilde{\chi}_{1 , {\bf -k}}
 (\eta) \rangle \right]  -\frac{1}{2} \right\}
 + \left( \tilde{\chi}_1 \rightarrow \tilde{\chi}_2 \, ; \, \Omega_{1 ,
 {\bf k}} (\eta) \rightarrow \Omega_{2 , {\bf k}} (\eta) \right) \, ,
\end{eqnarray}
where
\begin{equation}
 \Omega_{i , {\bf k}} (\eta) = {\cal W}^2_{i , {\bf k}}
 (\eta_0) a^2 (\eta)  + \frac{a'^{2} (\eta)}{a^2 (\eta)}
\end{equation}
and $\eta_0 \equiv \eta(t_0)$ is the initial conformal time.

The correlation functions can be expressed in terms of the mode 
functions as in Eq.~(\ref{chi-chifluc}):
\begin{eqnarray}
 \langle \tilde{\chi}'_{i , {\bf k}} (\eta) \tilde{\chi}'_{i ,
 {\bf -k}} (\eta) \rangle &=& | f_{i , {\bf k}}' (\eta) |^2 \, ,
 \nonumber \\
 \langle \tilde{\chi}_{i , {\bf k}} (\eta) \tilde{\chi}_{i ,
 {\bf -k}} (\eta) \rangle &=& | f_{i , {\bf k}} (\eta) |^2 \, .
\end{eqnarray}
In the weak coupling limit, and with appropriate choices for the
initial expectation values of the quantum fields, we may
approximate
\begin{equation}
 {\cal W}_{1 , {\bf k}} (t_0) \approx {\cal W}_
 {2 , {\bf k}} (t_0) \approx {\cal W}_{{\bf k}}
 (t_0) \, ; \,\,\,\, {\cal W}_{ {\bf k}}
 (t_0) = \left[ \frac{k^2}{ a^2 (t_0)} + m_{\Phi}^2
 \right]^{\frac{1}{2}} \, .
\end{equation}

This approximation simplifies the expression for the baryon number
density which is given by the current in Eq.~(\ref{Bcurrent}):
$\hat{n}_{\rm B}$~=~$i (\Phi^\dagger \dot{\Phi} - \dot{\Phi}^\dagger
\Phi)$.  In parallel with the discussion above for the particle
number density, we find
\begin{equation}
 \hat{n}_{\rm B}(t) = \frac{a^3 (t_0)}{2} \int
 \frac{d^3 {\bf k}}{(2 \pi)^3} \, i \, \left[
 \Phi^{\dagger}_{\bf -k} (t) \dot{\Phi}_{\bf k} (t)
 + \dot{\Phi}_{\bf -k} (t) \Phi^{\dagger}_{\bf k} (t)
 - \dot{\Phi}^{\dagger}_{\bf -k} (t) \Phi_{\bf k} (t) -
 \Phi_{\bf -k} (t) \dot{\Phi}^{\dagger}_{\bf k} (t)
 \right] \, .
\end{equation}
After converting to conformal time, the expectation value of the
baryon number density is found to be
\begin{eqnarray}
 n_{\rm B} (\eta) \equiv \langle \hat{n}_{\rm B} \rangle (\eta) &=&
 \frac{a^3 (\eta_0)} {a^3 (\eta)} \left( \chi^0_2 (\eta) \chi^{0 \,'}_1
 (\eta) - \chi^0_1 (\eta) \chi^{0 \,'}_2 (\eta) \right)
 \nonumber \\
 && + \frac{a^3 (\eta_0)}{a^3 (\eta)} \int \frac{d^3 {\bf k}}
 {(2 \pi)^3} \left[\langle \tilde{\chi}_{1 , {\bf k}}' (\eta)
 \tilde{\chi}_{2 , {\bf -k}} (\eta) \rangle - \langle
 \tilde{\chi}_{1 , {\bf k}} (\eta) \tilde{\chi}_{2 , {\bf -k}}'
 (\eta) \rangle \right] \, .
\end{eqnarray}
Like the particle number density, it can be separated into a mean
value plus its quantum fluctuations.  Just as in the earlier
computation for the term $\langle \tilde{\chi}_1 \tilde{\chi}_2
\rangle$, the quantum fluctuations can be computed by introducing a
vertex $6i \lambda \left[ (\chi^0_1)^2 - (\chi^0_2)^2 \right]
\tilde{\chi}_1 \tilde{\chi}_2$ at the lowest order, which yields
\begin{eqnarray}
 n_{\rm B} (\eta) &=& \frac{a^3 (\eta_0)}
 {a^3 (\eta) } \left( \chi^0_2 (\eta) \chi^{0 \,'}_1 (\eta) -
 \chi^0_1 (\eta) \chi^{0 \,'}_2 (\eta) \right)
 \nonumber \\
 && + 6 \, i\,a^3 (\eta_0)\, \lambda \, \int_{\eta_0}^{\eta} \,
 d\eta' \, \left\{ \left[ \chi_1^0 (\eta') \right]^2 - \left[
 \chi_2^0 (\eta') \right]^2 \right\} \left\{ \left[ f'_{1 ,
 {\bf k}} (\eta) f^{*}_{1 , {\bf k}} (\eta') f_{2 , {\bf k}}
 (\eta ) f^{*}_{2 , {\bf k}} (\eta') - {\rm c. c.} \right] \right.
 \nonumber \\
 && \left.
 \,\,\,\,\,\,\,\,\,\,\,\,\,\,\,\,\,\,\,\,\,\,\,\,\,\,\,\,\,\,\,\,\,
 \,\,\,\,\,\,\,\,\,\,\,\,\,\,\,\,\,\,\,\,\,\,\,\,\,\,\,\,\,\,\,\,\,
 \,\,\,\,\,\,\,\,\,\,\,\, - \left[ f_{1 , {\bf k}} (\eta)
 f^{*}_{1 , {\bf k}} (\eta') f'_{2 , {\bf k}} (\eta ) f^{*}_{2 ,
 {\bf k}} (\eta') - {\rm c. c.} \right] \right\} \, .
\end{eqnarray}

The baryon number density can be evaluated once the mode functions
and field expectation values are determined by the method described
in Section~IV.  In order to ensure the canonical quantization rules,
the mode functions must satisfy the relation
\begin{equation}
 f'_{i , {\bf k}} (\eta) f^{*}_{i , {\bf k}} (\eta) -
 f_{i , {\bf k}} (\eta) f_{i , {\bf k}}^{* '} (\eta) = - i \, ,
\end{equation}
which follows from the mode equations~(\ref{modeeq}).  With the
initial states described by the adiabatic modes for uncoupled
harmonic oscillations, the mode functions have the following
initial values at the conformal time $\eta_0$:
\begin{equation}
 f_{i , {\bf k}} (\eta_0) =\frac{ a(\eta_0)}{ {\sqrt{ 2 \, a^3
 (\eta_0) \, {\cal W}_{ {\bf k}}  (\eta_0)}} } \, , \,\,\,\,\,
 f'_{i , {\bf k}} (\eta_0) =\frac{ a'(\eta_0)}{ {\sqrt{ 2 \, a^3
 (\eta_0) \, {\cal W}_{ {\bf k}}  (\eta_0)}} } -i \frac{ a^2
 (\eta_0)  \, {\cal W}_{ {\bf k}} (\eta_0)  }{ {\sqrt{ 2 \, a^3
 (\eta_0) \, {\cal W}_{ {\bf k}} (\eta_0)}} } \,.
\label{initialcondmode}
\end{equation}
As expected, these initial conditions guarantee that the baryon
and particle number densities have zero quantum fluctuations
initially.

Consistent with the initial conditions~(\ref{initialcond}), the
initial expectation values of the conformally rescaled fields are
\begin{equation}
 \chi_1^0 (\eta_0) = 0\,\,\, ,\,\,\, \chi_1^{0 '} (\eta_0) = 0
 \,\,\, ; \,\,\,\,\,  \chi_2^0 (\eta_0) = \sqrt{2} \, a(\eta_0) \,
 \Phi_0 \,\,\, ,  \,\,\, \chi_2^{0 '} (\eta_0) = \sqrt{2} \,
 a'(\eta_0) \, \Phi_0 \, .
\label{initialcondmean}
\end{equation}
Together with (\ref{initialcondmode}), these initial conditions
guarantee that the initial baryon asymmetry is zero: $n_{\rm B}
(\eta_0) = 0$.

In the next section, we will present the results of our calculation
that assumes a radiation-dominated epoch in which the scale factor
follows $a(\eta) = m_\Phi \eta$.  We find that the results are
similar in a matter-dominated universe and will not present explicit
results for this case.

\section{Results}

Before we describe the numerical solutions to the coupled
equations, let us try to understand the issue of ultraviolet
divergence associated with the loop integrals in these
equations. The large-$k$ behavior of the mode functions can
be obtained from WKB type solutions to the mode equations:
\begin{equation}
 f_{i , {\bf k}} (\eta) = \frac{a(\eta_0)}{{2 \sqrt{2 \, a^3
 (\eta_0) \, {\cal W}_{{\bf k}} (\eta_0)}}} \left[(1 + \gamma)
 {\cal D}^{*}_{i, {\bf k}} (\eta) +  (1 - \gamma) {\cal D}_
 {i, {\bf k}} (\eta) \right] \, ,
\label{f-largek}
\end{equation}
where
\begin{equation}
 {\cal D}_{i, {\bf k}} (\eta) = e^{ \int^{\eta}_{\eta_0}
 {\cal R}_{i, {\bf k}} (\eta') d\eta' }
\label{d-largek}
\end{equation}
and
\begin{equation}
 {\cal R}_{i, {\bf k}} (\eta) = - i k -\frac{i}{2 k}
 {\cal M}^2_{\chi_i} (\eta) - \frac{1}{4 k^2} \frac{d}{d \eta}
 {\cal M}^2_{\chi_i} (\eta) + {\cal O} \left( \frac{1}{k^3}
 \right) + \cdot \cdot \cdot \, .
\label{r-largek}
\end{equation}
The parameter $\gamma $ that appears in Eq.~(\ref{f-largek}) can
be determined from the initial conditions~(\ref{initialcondmode})
on the mode functions.  It has the large-$k$ expansion
\begin{equation} \gamma = -1
 - \frac{i}{k} \left( \frac{a' (\eta_0)}{ a( \eta_0)} \right)+
 {\cal O} \left( \frac{1}{k^3} \right) +
 \cdot\cdot\cdot \, ,
\end{equation}
which secures
\begin{equation}
 |f_{i , {\bf k}} (\eta)|^2 = \frac{1}{2 k} - \frac{1}{4 k^3}
 \left( {\cal M}^2_{\chi_i} (\eta) - \frac{a^{' 2} (\eta_0)}
 {a^2 (\eta_0)} \right) + {\cal O} \left(\frac{1}{k^4}\right) +
 \cdot\cdot\cdot \, .
\end{equation}
These large-$k$ behaviors lead to quadratic and logarithmic
divergence for the loop integrals as follows:
\begin{equation}
 \langle \tilde{\chi}_i^2 \rangle (\eta) = \frac{\Lambda^2 }{8
 \pi^2} - \frac{1}{8 \pi^2} \left( {\cal M}^2_{\chi_i} (\eta) -
 \frac{a^{'2} (\eta_0)}{a^2 (\eta_0)} \right) \ln \frac{\Lambda}
 {\kappa} + {\rm finite } \, ,
\end{equation}
where $\Lambda$ is the ultraviolet momentum cutoff and $\kappa$
is the renormalization scale.  Let us recall that the Affleck-Dine
model under consideration is used as an effective field theory to
describe the relevant low-energy physics from a supersymmetric 
grand unified theory.  $\Lambda$ is therefore a physical cutoff 
which can be set at the scale of grand unification, $\Lambda 
\approx M_G $, while $\kappa$ can be chosen to be the supersymmetry 
breaking scale, $\kappa \approx M_S \approx m_{\Phi}$.  Since the 
relevant terms involved in the equations for the field expectation 
values are of the form $ \left(\langle \tilde{\chi}^2_1 \rangle - 
\langle\tilde{\chi}^2_2 \rangle \right) $, only the logarithmic 
dependence on the cutoff $\Lambda$ remains.  For weak couplings, 
such a mild cutoff dependence does not present any difficulties for 
our numerical study.  Using Eqs.~(\ref{f-largek}), (\ref{d-largek}), 
and (\ref{r-largek}), the quantum fluctuations $\langle \tilde{\chi}_1
\tilde{\chi}_2 \rangle $ prove to be ultraviolet safe.

\begin{figure}[tp]
\begin{center}
\leavevmode
\epsfxsize=5in
\epsffile{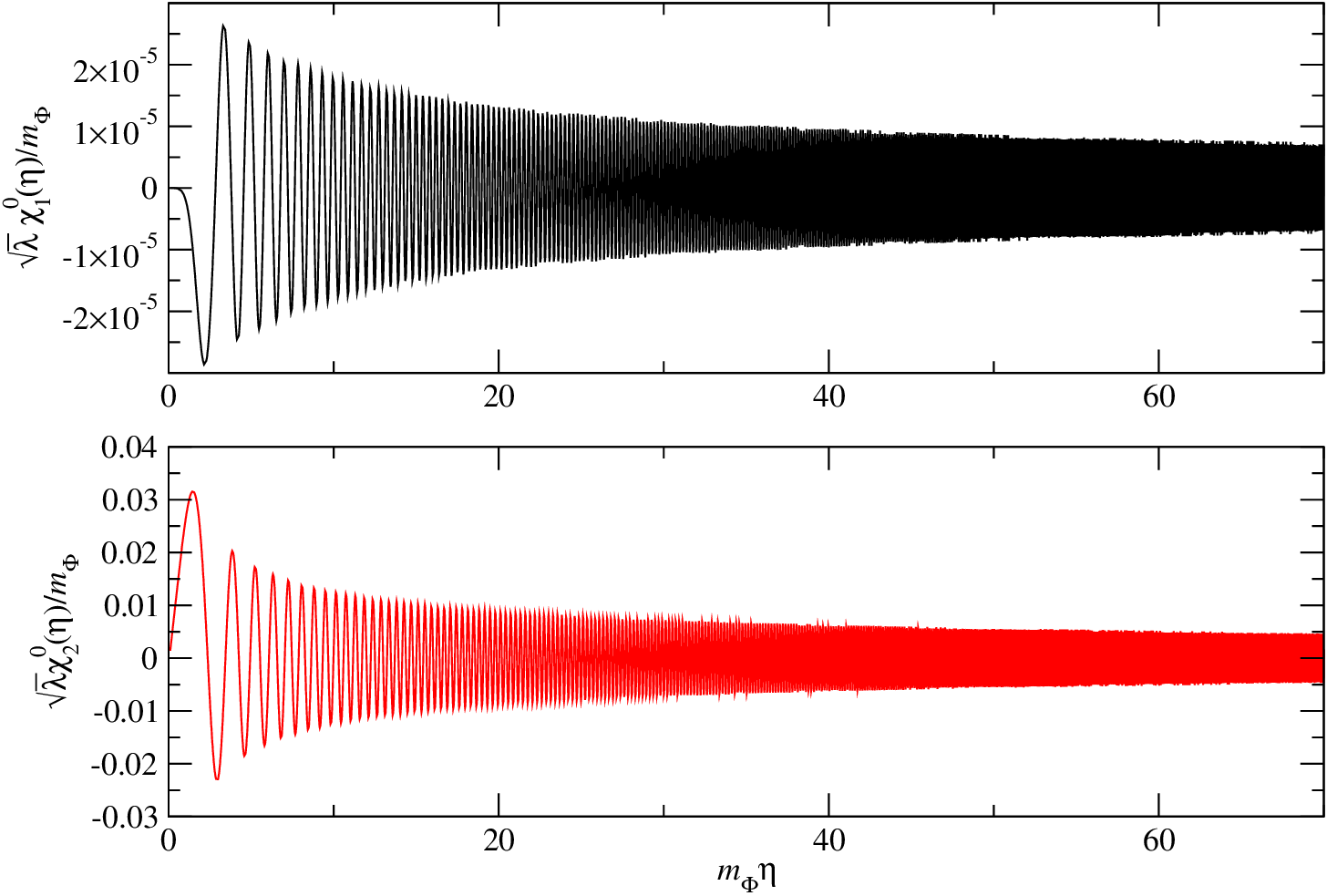}
\caption{Time evolution of the field expectation values $\chi_1^0$
and $\chi_2^0$ for $\phi_0 \equiv{\sqrt\lambda}\Phi_0/m_\Phi = 10^{-2}$
and $\lambda = 10^{-3}$.}
\label{fig1}
\end{center}
\end{figure}

\begin{figure}[htbp]
\begin{center}
\leavevmode
\epsfxsize=5in
\epsffile{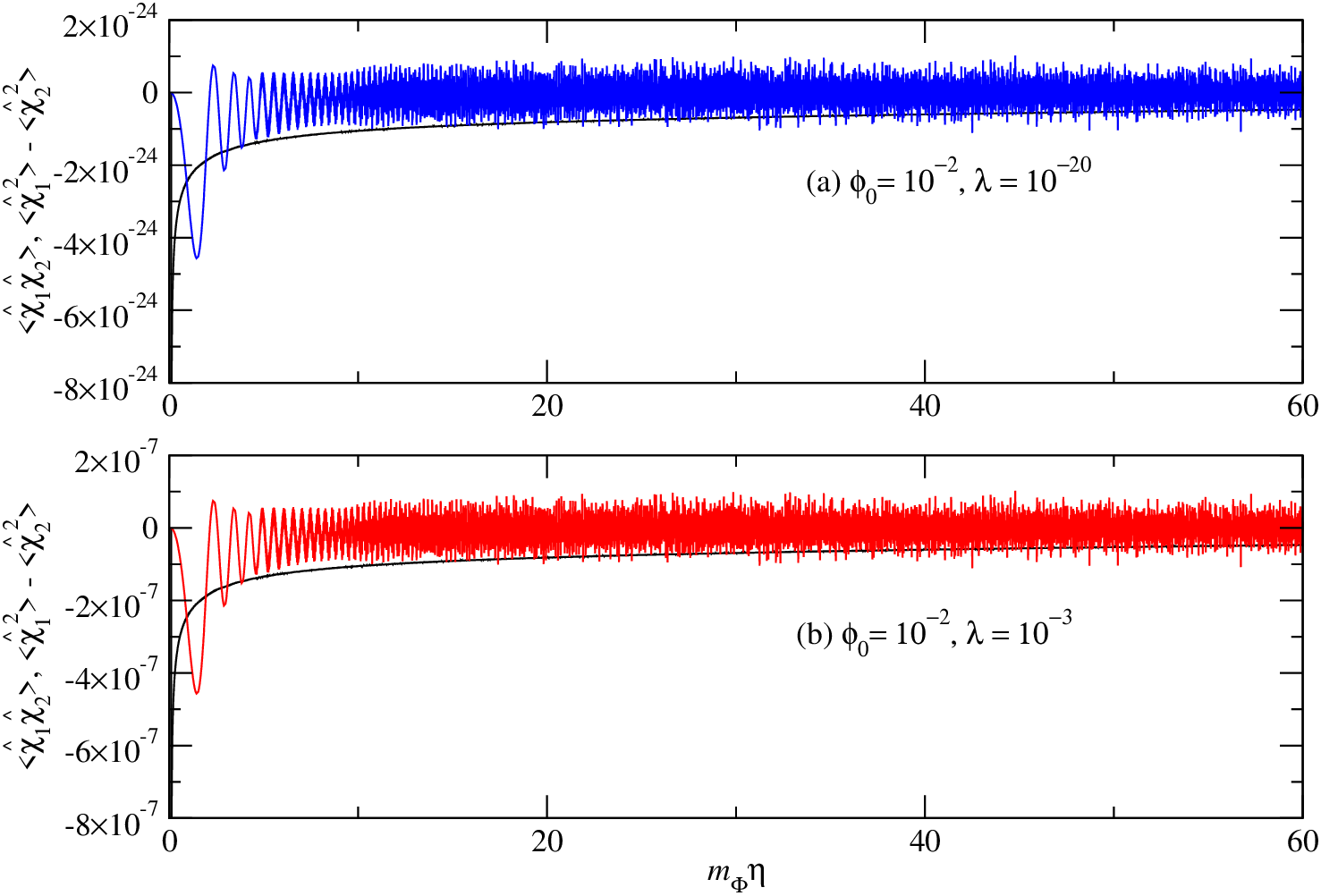}
\caption{Time evolution of $\langle \tilde{\chi}_1^2 \rangle -
\langle \tilde{\chi}_2^2\rangle$ (smooth curves) and
$\langle \tilde{\chi}_1 \tilde{\chi}_2 \rangle$ (oscillating curves)
for $\phi_0=10^{-2}$ and (a) $\lambda=10^{-20}$ and (b) $\lambda=
10^{-3}$.  Note that we have used the rescaled fields
$\hat{\chi}_i\equiv {\sqrt\lambda}\tilde{\chi_i}/m_\Phi$.}
\label{fig2}
\end{center}
\end{figure}

\begin{figure}[htbp]
\begin{center}
\leavevmode
\epsfxsize=5in
\epsffile{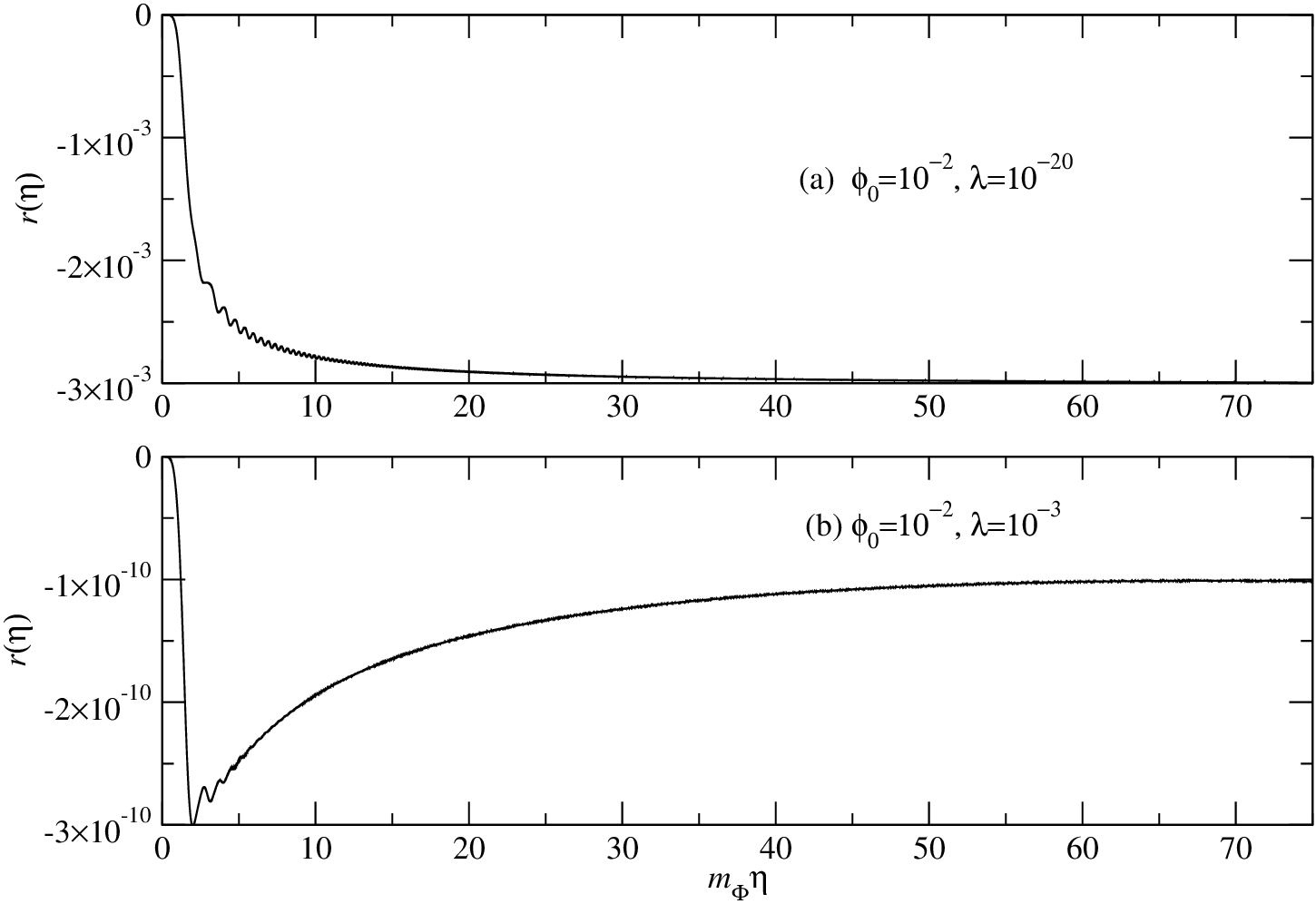}
\caption{Time evolution of the ratio of the baryon to particle
number density, $r(\eta) = n_{\rm B} (\eta) / n (\eta)$, for
$\phi_0=10^{-2}$ and (a) $\lambda = 10^{-20}$ and (b) $\lambda =
10^{-3}$.}
\label{fig3}
\end{center}
\end{figure}

\begin{figure}[htbp]
\begin{center}
\leavevmode
\epsfxsize=5in
\epsffile{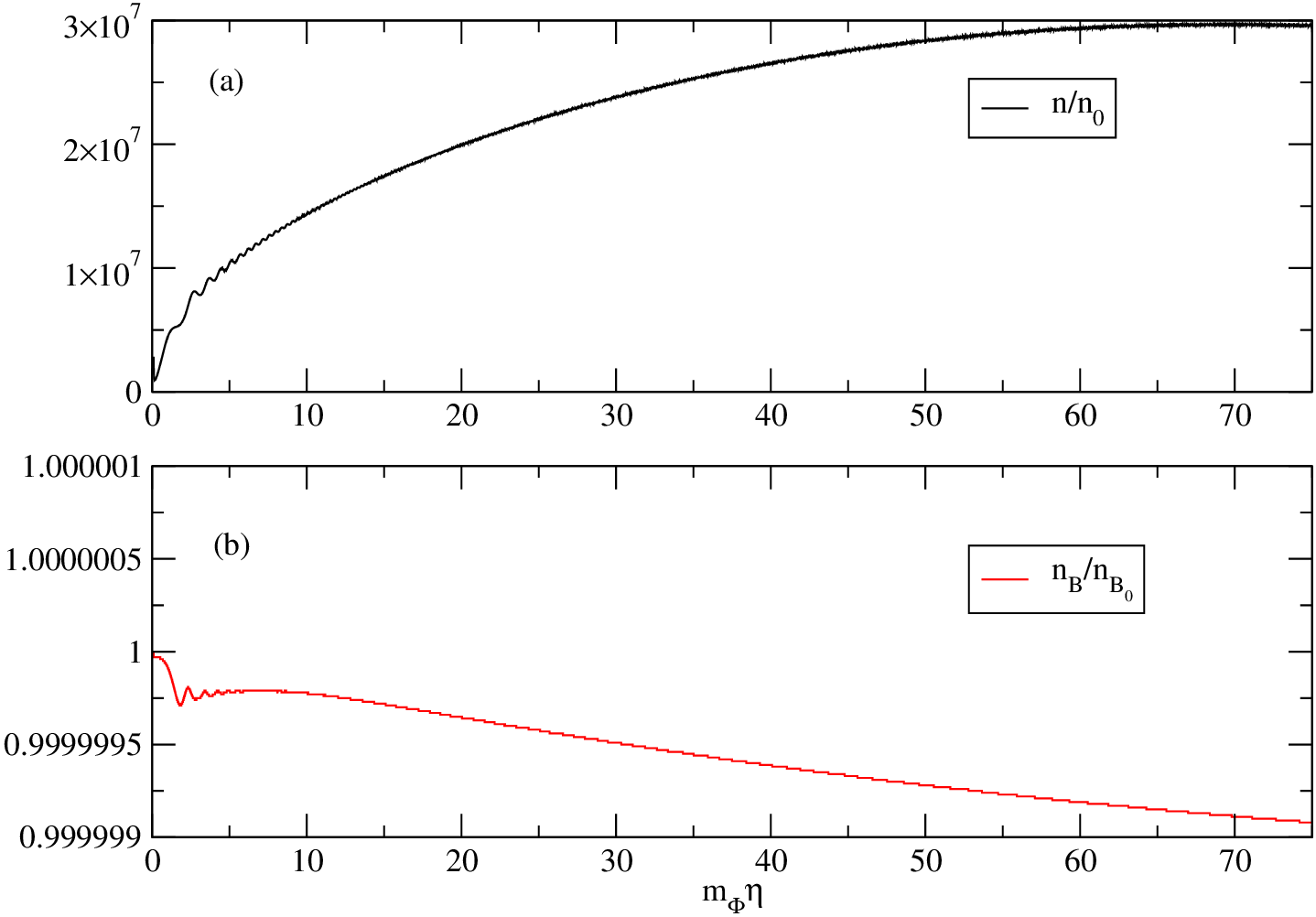}
\caption{(a) Time evolution of the ratio of the total particle
number density to its classical value for $\phi_0=10^{-2}$ and
$\lambda=10^{-3}$. (b) Same as (a) but for the baryon number
density. }
\label{fig4}
\end{center}
\end{figure}

\begin{figure}[htbp]
\begin{center}
\leavevmode
\epsfxsize=5in
\epsffile{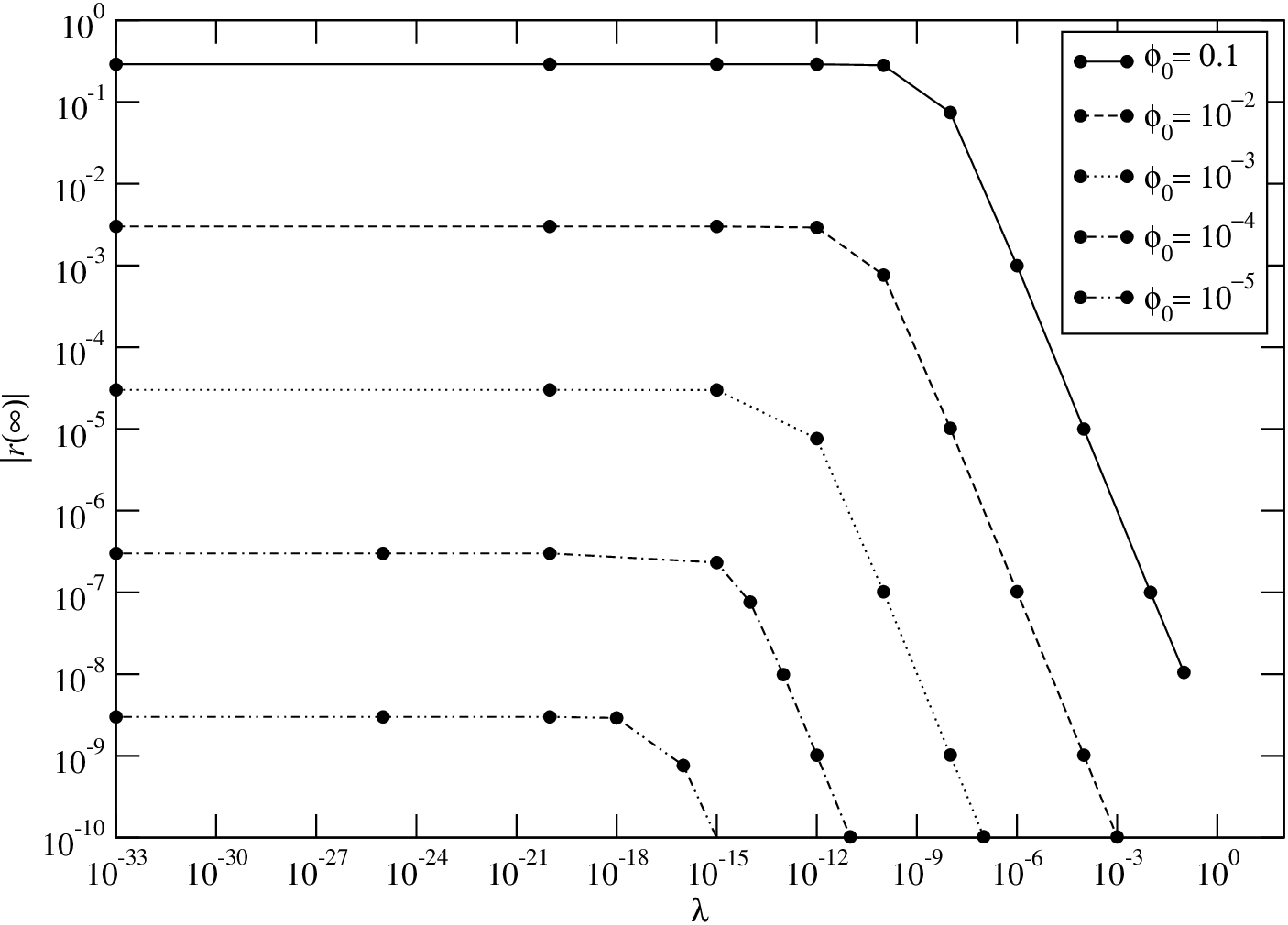}
\caption{Ratio of the baryon to particle number density at large
times for various initial values of $\Phi_0$ and couplings
$\lambda$.}
\label{fig5}
\end{center}
\end{figure}

Let us turn now to the numerical results of our study.  
Figure~1 shows the evolution of the expectation values of the
quantum fields $ \chi^0_1 ( \eta)$ and $\chi^0_2 ( \eta) $ in
units of $m_\Phi/\sqrt{\lambda}$, the scale of which is set by
the dimensionless parameter $\phi_0\equiv{\sqrt\lambda}\Phi_0/m_\Phi
=10^{-2}$.  According to Eqs.~(\ref{chi1eq}) and (\ref{chi2eq}),
the evolution of the field expectation values are determined by
the fluctuations $(\langle \tilde{\chi}_1^2 \rangle - \langle
\tilde{\chi}_2^2 \rangle )(\eta)$ and $\langle \tilde{\chi}_1
\tilde{\chi}_2 \rangle (\eta)$, which are displayed in Fig.~2.

For an extremely weak coupling, e.g., $\lambda = 10^{-20}$,
Fig.~2(a) shows that the quantum fluctuations are very small and
$\chi^0_1 (\eta)$ and $\chi^0_2 (\eta)$ follow essentially their
classical trajectories which correspond to oscillations with
time-dependent frequencies.  For a larger, but still weak, coupling
(e.g., $\lambda = 10^{-3}$), the fluctuations become more significant
(see Fig.~2(b)) and the amplitudes of the oscillations of $\chi^0_1$
and $\chi^0_2$ become damped, as can be seen from Fig.~1.  From
Eq.~(\ref{chi1eq}) and the initial conditions (\ref{initialcondmean}),
we can estimate the order of magnitude of $\chi_1^0 (\eta)$ to be
\begin{equation}
 \chi_1^0 (\eta) \approx \lambda \frac{\left[\chi_2^0 (\eta)
 \right]^3}{m^2_{\Phi}}\, , \label{chi1approx}
\end{equation}
which is consistent with Fig.~1.

The time evolution of the baryon asymmetry, as measured by the
ratio of the baryon number density to the  particle number density,
$r(\eta) = n_{\rm B} (\eta) /n (\eta)$, is shown in Fig.~3.  As
stated earlier, the initial baryon asymmetry is zero due to the
initial conditions in Eqs.~(\ref{initialcondmode}) and
(\ref{initialcondmean}).  For very weak couplings, the evolution
of $r$ is determined by the classical dynamics and is depicted in
Fig.~3(a).  It starts from zero and decreases toward an equilibrium
value $r (\infty) \approx - 10 \phi_0^2$.  This saturation value
can be understood as follows:
\begin{equation}
 |r (\infty)| = \frac{|n_{\rm B} (\infty)|}{n (\infty)} \approx
 \frac{ \chi_1^0 \chi_2^0}{\left(\chi_2^0 \right)^2 } \approx
 \left[ \frac{\lambda \, \left(\chi^0_2 \right)^4}{m_{\Phi}^2}
 \right] \left[ \frac{1}{\left(\chi_2^0 \right)^2} \right]
 \approx \frac{\lambda \, \Phi_0^2}{m_{\Phi}^2} \, ,
\end{equation}
where the approximation for $\chi^0_1$ in Eq.~(\ref{chi1approx})
has been applied to compute the baryon number density $n_{\rm B}$,
but neglected in the number density $n$ as compared with that of
$\chi^0_2$.  The negative baryon asymmetry should not be a
concern as the sign can be fixed by the baryon number assignment
for the field $\Phi$.  The alert reader will recognize that the
estimate of $r (\infty)$ above is nothing but the result quoted
in Eq.~(\ref{rAD}).

As can be seen from Fig.~3(b), an interesting phenomenon occurs
for larger couplings when the quantum fluctuations become large,
typically of the order of the field expectation values. The
growth of quantum fluctuations is a consequence of parametric
amplification~\cite{lee,linde} induced by the oscillating mean
fields.  In this case the part of the ratio $r(\eta)$ generated
from the quantum fluctuations turns out to dilute the mean-field
contribution, and therefore drives the ratio toward a much smaller
equilibrium value: $|r(\infty)| \approx 10^{-5} \lambda \Phi_0^4/
m_{\Phi}^4$.  It thus appears that, under appropriate conditions,
the effects of the nonequilibrium quantum fluctuations can wash out
the baryon asymmetry generated from the classical dynamics.  The
reason for this may be understood as follows.  The set of coupled
equations, Eqs.~(\ref{modeeq}), (\ref{chi1eq}), (\ref{chi2eq}),
(\ref{chi-chifluc}) and (\ref{chi1-chi2fluc}), which we solve to
obtain the baryon asymmetry, involves fluctuation terms with
nonzero baryon number such as $(\langle \tilde{\chi}_1^2 \rangle
- \langle \tilde{\chi}_2^2 \rangle)$ and $\langle \tilde{\chi}_1
\tilde{\chi}_2 \rangle$.  As such, the amount of baryon
asymmetry that can be generated is limited by the Hartree
approximation implemented in these equations.  On the other hand,
as a result of the nature of the couplings under consideration,
it is possible to create a large particle number density from the
fluctuation term $(\langle \tilde{\chi}_1^2 \rangle + \langle
\tilde{\chi}_2^2 \rangle)$, which has nonzero particle number but
zero baryon number under the global $U(1)$ symmetry.  This can be
seen in Fig.~4, which shows that for a sufficiently large coupling
the total particle number density deviates substantially from its
classical value, while the baryon number density remains close to
its classical value.  A similar plot for a much weaker coupling,
e.g., $\lambda = 10^{-20}$, would show that both the particle
number density and the baryon number density remain essentially
at their respective classical values.  We are thus led to the
conclusion that the ratio $r$ can be reduced significantly through
nonequilibrium quantum fluctuations with a proper choice of the
couplings.

Figure~5 shows the saturation values of $r$ as a function of the
coupling $\lambda$ for a wide range of values of $\Phi_0$.  Plotted
are equal $\phi_0$ contours, where $\phi_0 = {\sqrt\lambda}\Phi_0/
m_\Phi$.  For a given contour, the saturation value for small
values of $\lambda$ is given by the classical result, $|r(\infty)|
\approx 10\phi_0^2$, which corresponds to the horizontal part of
the curve.  When $\lambda$ becomes larger than about $10^{-6}\phi_0^2$,
the nonequilibrium effects start to operate and significantly diminish
the saturation ratio.  For a wide range of parameters, this can
result in $r$ saturating toward the observed value of the baryon
asymmetry, as shown by the right end-points of the contours.

As an illustration, let us recall the sample values of the model
parameters cited after Eq.~(\ref{rAD}), which corresponds to the
left end of the $\phi_0=10^{-2}$ contour in Fig.~5.  Now let us
entertain the possibility that the supersymmetry breaking scale may
be as high as the grand unification scale, i.e., $M_S \alt M_G$.  We
then find $\lambda \alt 10^{-3}$ and the baryon asymmetry is reduced
by the nonequilibrium dynamics to the present observed value,
$|r(\infty)| \approx 10^{-10}$, corresponding to the right end of
the $\phi_0=10^{-2}$ contour in Fig.~5.

\section{Conclusions}

We have studied the effects of nonequilibrium dynamics in the
Affleck-Dine mechanism of baryogensis.  Within the context of
the model (\ref{model}), we find that, for very weak couplings, the
ratio of the baryon to particle number density is approximately
$10 \lambda \Phi_0^2/ m^2_{\Phi}$, as dictated by the classical
dynamics, but for larger couplings it is driven by the nonequilibrium
effects toward a smaller value given by $|r| \approx 10^{-5} \lambda
\Phi_0^4/ m_{\Phi}^4$.  Interestingly, in some cases these
nonequilibrium effects can reduce the baryon asymmetry generated in
the classical Affleck-Dine mechansim to the present observed value
without having recourse to additional dilution processes.  This
encouraging result suggests that nonequilibrium dynamics can play
a significant role in the evolution of the early Universe.  It is
therefore important to explore other possible manifestations of
these nonequilibrium effects.  The methodology developed and
presented in this paper will be useful for such studies.

\begin{acknowledgments}
This work was supported in part by the National Science Council,
ROC, under grant NSC92-2112-M-001-029 (NSC93-2112-M-259-007) as
well as the short-term visiting program of the Royal Society, 
and in part by the U. S. Department of Energy under grant
DE-FG02-84ER40163.  CNL would like to thank the members of the
Institute of Physics at the Academia Sinica in Taipei, Taiwan,
especially H.-Y. Cheng, C.-Y. Cheung, S.-P. Li, and K.-W. Ng, for
their hospitality.  He also thanks S.-Y. Wang for a useful 
discussion.
\end{acknowledgments}

\end{document}